\documentclass[showpacs,amsmath,amssymb]{revtex4}
\usepackage{graphicx}
\usepackage{dcolumn}
\usepackage{bm}
\usepackage{epsfig}

\newcommand{\g}{\gamma}

\newcommand{\jpsi}{J/\psi}

\newcommand{\ppp}{\pi^+\pi^- \pi^{0}}
\newcommand{\pip}{\pi^+}

\newcommand{\pin}{\pi^-}
\newcommand{\pio}{\pi^0}

\newcommand{\ks}{K^0_S}

\newcommand{\pp}{\pi^+\pi^-}

\newcommand{\pppp}{\pi^+\pi^-\pi^+\pi^-}

\newcommand{\gpppp}{\gamma \pi^+\pi^-\pi^+\pi^-}

\newcommand{\gksks}{\gamma K^0_s K^0_s}

\newcommand{\gff}{\gamma f_{2} f_{2}}
\newcommand{\gfmfm}{\gamma f_{2}(1270) f_{2}(1270)}
\newcommand{\grr}{\gamma \rho^{0}\rho^{0}}
\newcommand{\ff}{f_{2} f_{2}}
\newcommand{\ppppp}{\pi^0 \pi^+\pi^-\pi^+\pi^-}

\newcommand{\omegpp}{\omega \pi^+\pi^-}
\newcommand{\omegf}{\omega f_{2}}

\newcommand{\rpf}{f_{2} \rho^{\pm}\pi^{\mp}}

\newcommand{\etac}{\eta_{c}}

\newcommand{\ar}{\rightarrow}

\newcommand{\mpppp}{M_{\pi^+\pi^-\pi^+\pi^-}}
\newcommand{\mpp}{M_{\pi^+\pi^-}}

\newcommand{\bfg}{\begin{figure}[htpb]}
\newcommand{\efg}{\end{figure}}
\newcommand{\bitm}{\begin{itemize}}
\newcommand{\eitm}{\end{itemize}}
\newcommand{\bnum}{\begin{enumerate}}
\newcommand{\enum}{\end{enumerate}}
\newcommand{\btbl}{\begin{table}[htp]}
\newcommand{\etbl}{\end{table}}
\newcommand{\btbu}{\begin{tabular}[htp]}
\newcommand{\etbu}{\end{tabular}}
\newcommand{\bcl}{\begin{center}}
\newcommand{\ecl}{\end{center}}

\newcommand{\beq}{\begin{equation}}
\newcommand{\eeq}{\end{equation}}
\newcommand{\beqr}{\begin{eqnarray}}
\newcommand{\eeqr}{\end{eqnarray}}

\usepackage{epsfig}
\renewcommand{\baselinestretch}{1.0}
\begin{document}
\normalsize

\parindent = 0.5 in

\renewcommand{\baselinestretch}{1.0}
\textheight=22cm

\title{\large \bf \boldmath First Measurement of the Branching Ratio of
  $\jpsi\ar \gfmfm$ }

\vskip 0.5cm

\author{
M.~Ablikim$^{1}$, J.~Z.~Bai$^{1}$, Y.~Ban$^{10}$,
J.~G.~Bian$^{1}$, X.~Cai$^{1}$, J.~F.~Chang$^{1}$,
H.~F.~Chen$^{16}$, H.~S.~Chen$^{1}$, H.~X.~Chen$^{1}$,
J.~C.~Chen$^{1}$
% ...
\\(BES Collaboration)\\
\vspace{0.2cm}
$^1$ Institute of High Energy Physics, Beijing 100039, People's Republic of China\\
% ...
$^{10}$ Peking University, Beijing 100871, People's Republic of China\\
% ...
$^{16}$ University of Science and Technology of China, Hefei 230026, People's Republic of China\\
% ...
}

\begin{abstract}
Using 58 million $\jpsi$ events taken
with the  BES\,II detector at the Beijing Electron Positron
Collider, a new decay mode $\jpsi\ar \gfmfm \ar \gpppp$ is
observed for the first time. The branching ratio is determined to be
 $Br(\jpsi\ar\gff)=(9.5\pm0.7\pm1.6)\times 10^{-4}$, where
the quoted errors are statistical and systematic, respectively.
\end{abstract}


\maketitle

\section{Introduction}

Precise branching ratios are needed in order to better understand
$\jpsi$ physics.  Unfortunately, only about 50 to 60\% of $\jpsi$
decay modes have been observed so far~\cite{pdg}. A sample of 58
million $\jpsi$ events has been accumulated with the upgraded Beijing
Spectrometer (BES\,II). With this sample, the world's largest, it is
possible to systematically study $\jpsi$ decays.

For the decay $\jpsi \ar \gpppp$, studies have been done by MARK\,III
~\cite{mark3,Adler:1988kg}, DM2 ~\cite{dm2} and BES
~{\cite{Bai:1999mm,besetac}.  A large part of the final state is
from $\jpsi \ar \grr$ ~\cite{mark3}.  MARK\,III also reported the
observation of $\jpsi\ar \gamma \etac \ar \gff\ar\gpppp$
~{\cite{Adler:1988kg}.  In this paper, evidence for the decay
$\jpsi\ar \gfmfm\ar \gpppp$ is observed, and the branching ratio of
$\jpsi\ar \gfmfm$ is measured for the first time.


\section{BES Detector}

BES is a conventional solenoidal magnet
 detector~\cite{bes,bes2}.
 A 12-layer vertex chamber (VTC)
surrounding the beam pipe provides trigger and trajectory information. A
forty-layer main drift chamber (MDC), located radially outside the
VTC, provides trajectory and energy loss ($dE/dx$) information for
charged tracks over $85\%$ of the total solid angle with a
momentum resolution of $\sigma _p/p = 0.0178 \sqrt{1+p^2}$ ($p$ in
$\hbox{\rm~GeV}/c$) and a $dE/dx$ resolution for hadron tracks of
$\sim 8\%$. An array of 48 scintillation counters surrounding the
MDC measures the time-of-flight (TOF) of charged tracks with a
resolution of $\sim 200$ ps for hadrons.  Radially outside the TOF
system is a 12 radiation length, lead-gas barrel shower counter
(BSC).  This measures the energies of electrons and photons over
$\sim 80\%$ of the total solid angle with an energy resolution of
$\sigma_E/E=21\%/\sqrt{E}$ ($E$ in~GeV).  Outside the solenoidal
coil, which provides a 0.4~Tesla magnetic field over the tracking
volume, is an iron flux return that is instrumented with three
double layers of proportional counters that identify muons of momentum greater
than 0.5~GeV/c.

A Geant3 based Monte Carlo, SIMBES, which simulates the
detector response, including interactions of secondary particles
in the detector material, is used in this analysis.
Reasonable agreement between data and Monte Carlo simulation is
observed in various channels tested, including $e^+e^-\to(\gamma)
e^+ e^-$, $e^+e^-\to(\gamma)\mu\mu$, $J/\psi\to p\bar{p}$,
$J/\psi\to\rho\pi$ and $\psi(2S)\to\pi^+\pi^- J/\psi$,
$J/\psi\to l^+ l^-$.

\section{Event Selection}
First a $\jpsi \ar \gpppp$
sample is selected.  Events are
required to have four good charged tracks and one or more photon
candidates. A good track, reconstructed from hits in the MDC, must be
well fitted to a helix originating from the interaction point; have a
polar angle, $\theta$, with $|\cos\theta|<0.8$; and a transverse
momentum greater than 60 MeV.

The TOF and $dE/dx$ information are used for particle
identification. To identify pions, we define:
$$\chi_{TOF}(i)=\frac{TOF_{measured}-TOF_{expected-i}}{\sigma_{TOF-i}}$$
$$\chi_{dE/dx}(i)=\frac{PHMP_{measured}-PHMP_{expected-i}}{\sigma_{PHMP-i}} $$
$$Prob(i)=Prob(\chi^{2}_{TOF}(i)+\chi^{2}_{dE/dx}(i),2).$$
Here ${\it i}$ denotes $\pi$, $K$ and $p$. If the charged track has only
TOF information or only $dE/dx$ information, then $Prob(i)$ is
determined using that system only.  To identify a pion, it is required
that: $Prob(\pi)>Prob(K)$ and $Prob(\pi)>Prob(p)$. At least three
tracks must be identified as pions.

To reduce the number of spurious low energy photons produced by
secondary hadronic interactions, photon candidates must have a minimum
energy of 30 MeV and be outside a cone with a half-angle of
$15^{\circ}$ around any charged track.

To get higher momentum resolution and to remove backgrounds, events
are kinematically fitted to the $\jpsi \ar \gpppp$ hypothesis, looping
over all photon candidates. The fit with the highest probability is
selected, and the $\chi^2$ of the fit is required to be less than 20.

For $\jpsi\ar\gpppp$ decay, a major source of background is from
$\jpsi \ar \ppppp$. To remove events containing a $\pi^0$ when there
are multiple photons, $|m({\gamma 1}{\gamma
2})-m(\pi^{0})|>60{\rm~MeV}$ is required if
$\overrightarrow{P}_{miss}$ is in the plane of the two photons,
${\gamma 1}$ and ${\gamma 2}$, i.e. $|\widehat{P}_{miss}
\cdot(\widehat{\hat{r}_{\gamma 1}\times \hat{r}_{\gamma
2}})|<0.15$. Here $\widehat{P}_{miss}$ is the unit vector of the
missing momentum of all charged tracks; $\hat{r}_{\gamma 1}$ and
$\hat{r}_{\gamma 2}$ are unit vectors in the ${\gamma 1}$ and ${\gamma
2}$ directions, respectively; and $m({\gamma 1}{\gamma 2})$ is the
invariant mass of ${\gamma 1}$ and ${\gamma2}$.  Two additional
requirements, $\chi^{2}_{\gpppp}<\chi^{2}_{\gamma\gamma\pppp}$ and
$P^{2}_{t\gamma}<0.0015\rm~GeV^{2}$, are used to further remove $\jpsi
\ar \ppppp$ background.  $P_{t\gamma}$ is the transverse momentum of
the $\pppp$ system with respect to the photon.  Finally, the
requirement $|U_{miss}|=|E_{miss}-P_{miss}|<0.07$ GeV is used to
reject events with multiple photons and charged kaons; here,
$E_{miss}$ and $P_{miss}$ are, respectively, the missing energy and
missing momentum calculated using only the charged particles, which
are assumed to be pions.

There are other possible backgrounds such as $\jpsi\ar \omegpp$,
$\jpsi\ar \gksks$, and $\jpsi\ar \gamma K^0_s K^{\pm} \pi^{\mp} $.
The $\jpsi\ar \omegpp$ background is eliminated by the requirement
$|M_{\ppp}-M_{\omega}|>40{\rm~MeV}$, where the $\pio$ in $\ppppp$ is
associated to the missing momentum and energy determined using only
the charged tracks. To remove the $\jpsi\ar \gksks$ background, we
require $|M_{\pp}-M_{\ks}|>25{\rm~MeV}$ for both $\pi^+ \pi^-$
pairs. To remove the background from $\jpsi\ar \gamma K^0_s K^{\pm}
\pi^{\mp}$, events are rejected if $\chi^{2}(\jpsi\ar \gamma
K^{\pm}\pi^{\mp}\pip\pin)<\chi^{2}(\jpsi\ar \gpppp)$ when
$|M_{\pp}-M_{\ks}|<25{\rm~MeV}$.


Figure \ref{mpipi4}(a) shows the scatter plot of
$M_{\pi 1\pi 4}$ versus $M_{\pi 2\pi 3}$ for surviving events with
$M_{\pppp}\geq 2.0{\rm~GeV}$, where $\pi 1$ ($\pi 3$) and $\pi 2$
($\pi 4$) are the
$\pi^+$ and $\pi^-$ with the higher (lower) momentum. There is a clear signal near
($M_{\rho^0}$, $M_{\rho^0}$). Since according to Monte Carlo study
most events from $\jpsi\ar \gff$ have
$\pip\pin\pip\pin$ invariant mass greater than 2.6 GeV, the
scatter plot of $M_{\pi 1\pi 4}$ versus $M_{\pi 2\pi 3}$ with
$M_{\pppp}\geq 2.6{\rm~GeV}$ is shown in Fig. \ref{mpipi4}(b). An
obvious cluster near ($M_{f_{2}}$, $M_{f_{2}}$) is seen besides the
signal near ($M_{\rho^0}$, $M_{\rho^0}$).

There is another possible $M_{\pip\pin}$ versus $M_{\pip\pin}$
combination, which is $M_{\pi 1\pi 2}$ versus $M_{\pi 3\pi 4}$. Monte
Carlo studies of $\jpsi\ar \gff$ and $\jpsi\ar \g\rho^{0}\rho^{0}$
show that the $\ff$ and $\rho^{0}\rho^{0}$ signal almost always appears
in the $M_{\pi 1\pi 4}$ versus $M_{\pi 2\pi 3}$ combination. Therefore we
use the $M_{\pi 1\pi 4}$ versus $M_{\pi 2\pi 3}$ combination to study
$\jpsi\ar\gff$ decay.  The $M_{\pi 1\pi 4}$ and $M_{\pi 2\pi 3}$
distributions for $M_{\pppp}\geq 2.6{\rm~GeV}$ are shown in Fig.
\ref{mpipi4}(c) and Fig. \ref{mpipi4}(d) respectively.  Besides the
$\rho$ peak, a peak near
$M_{f_{2}}$ (1.275~GeV) can be clearly seen in both of these plots.

\begin{figure}[htbp]
\centerline{ \hbox{\psfig{file=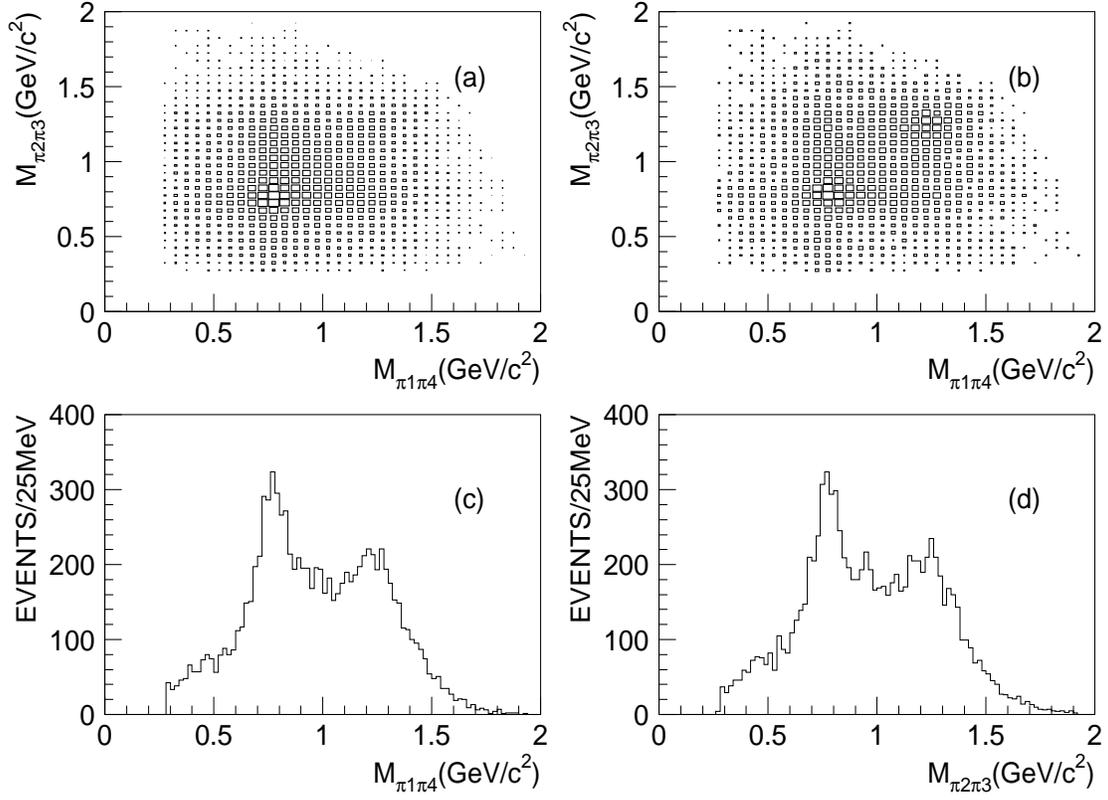,
  width=15cm, height=11cm} }}
\caption{(a) $M_{\pi 1\pi 4}$ versus $M_{\pi 2\pi 3}$ for
$\mpppp \geq 2.0{\rm~GeV}$. (b) $M_{\pi 1\pi 4}$
versus $M_{\pi 2\pi 3}$ for $\mpppp \geq 2.6{\rm~GeV}$. (c)
$M_{\pi 1\pi 4}$ and 
(d) $M_{\pi 2\pi 3}$ distributions with $\mpppp
\geq 2.6{\rm~GeV}$.}
\label{mpipi4}
\end{figure}

\section{Background Analysis }

Possible backgrounds are studied using simulated exclusive channels
normalized according to Particle Data Group (PDG) branching ratios, as
well as a 30 million event inclusive $\jpsi$ decay Monte Carlo sample
generated with Lundcharm ~\cite{chenjc}.
Simulated exclusive decays studied are $\jpsi\ar \grr$, $\jpsi\ar \gpppp$,
$\jpsi\ar \ppppp$, $\jpsi\ar \omegpp$, $\jpsi\ar\omegf\ar \omegpp$, and
$\jpsi\ar \pppp$, where events are generated according to
phase space and normalized by PDG ~\cite{pdg} branching ratios. After passing the
simulated events through our selection criteria, no
$f_{2}$ signal remains from these channels.

\begin{figure}[htbp]
\centerline{\hbox{\psfig{file=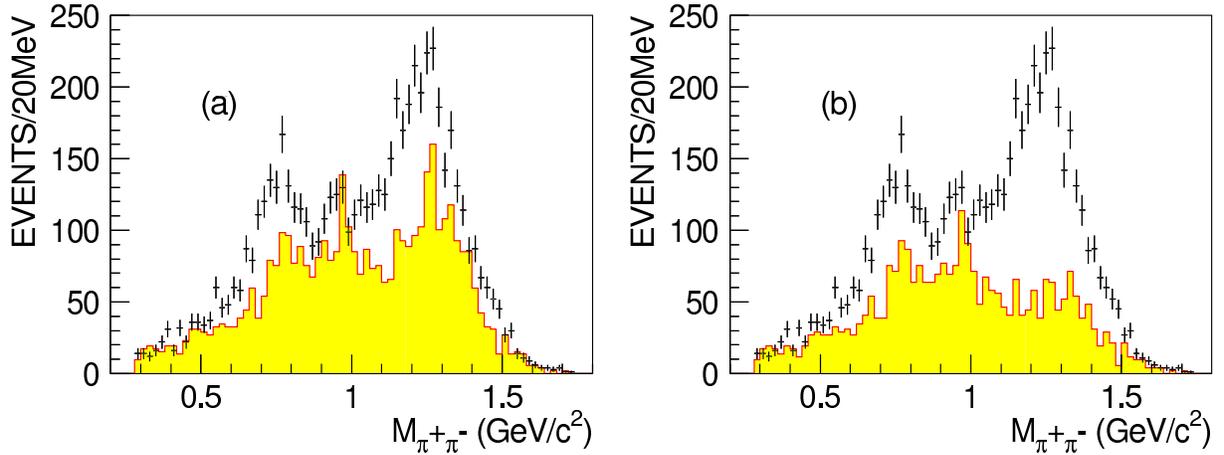, width=16cm, height=6cm}}}
\caption{The $\mpp$ spectra from data and
inclusive $\jpsi$ decay Monte Carlo data, which are
the sum of $M_{\pi 1\pi 4}$ and $M_{\pi 2\pi 3}$.
The dots with error bars are data, the shaded histogram is
from inclusive $\jpsi$ decay Monte Carlo data. (a) Including $\jpsi \ar \rpf$. (b) $\jpsi \ar \rpf$ removed.}
\label{mpipi30m}
\end{figure}

Fig.  \ref{mpipi30m} shows the $\mpp$ spectra
from data and from 30 million inclusive $\jpsi$ Monte Carlo decays
normalized to 58 million events, where
$\mpppp$ is required to be greater than 2.6~GeV and the invariant mass
of the other $\pi^+ \pi^-$ pair satisfies
${|M_{\pi^+\pi^-}-M_{f_{2}}|\leq 0.185{\rm~GeV}}$. This $\mpp$
spectrum is used to fit the $\jpsi \ar \gff$ signal in Section
\ref{results}. In Fig. \ref{mpipi30m}(a), there is an $f_{2}$ signal
from $\jpsi\ar \rpf$ decays from the inclusive $\jpsi$ Monte Carlo sample. After
removing this channel, there is no $f_{2}$ signal left in the
inclusive $\jpsi$ decay Monte Carlo events, as shown in
Fig. \ref{mpipi30m}(b). So the $f_{2}$ signal in the $\mpp$ spectrum
of inclusive $\jpsi$ Monte Carlo decays is caused by $\jpsi\ar \rpf$.

Although $\jpsi\ar \rpf$ decay is not listed in the PDG, we observe
this decay in $\jpsi \ar \ppppp$, and it gives a
background in the $\mpp$ spectrum of $\jpsi \ar \gff$. There is
an $f_{2}$ signal in the $\mpp$ distribution of $\jpsi \ar \ppppp$
when a $\rho^{\pm}$ mass requirement is made on the other
$\pi^{\pm}\pi^0$ pair. The number of fitted $f_{2}$ events in this $\mpp$
distribution is 2111, and the efficiency, obtained by a Monte Carlo
simulation, for this $\jpsi \ar \rpf$ selection is 0.67\%. The number
of $\jpsi \ar \rpf \ar \ppppp$ events is $N_{\rpf}=3.15\times
10^{5}$, which is used to determine the amount of $\rpf$
background in the $\mpp$ spectrum of $\jpsi \ar \gff$.


\section{Results and Systematic Errors}
\label{results}

The $\mpp$ spectrum for events with $\mpppp\ge2.6{\rm~GeV}$ and with 
one $\pp$ pair satisfying ${|\mpp-M_{f_{2}}|
\leq 0.185{\rm~GeV}}$ is
used to obtain the number of $\jpsi \ar \gff$ events.
The $\pp$ mass
spectrum is fitted with a $\rho^{0}$, a $f_{2}(1270)$, and
a 2nd order polynomial background plus $\rpf$
background using a binned maximum likelihood fit. The mass and width
of the $\rho^{0}$ are fixed at the PDG
values. The shape of the $f_{2}$ is determined by Monte Carlo simulation for $\jpsi \ar \gff$ ($M_{f_{2}}=1.275
\rm~GeV$, $\Gamma_{f_{2}}=0.185 \rm~GeV$).
The shape of the $\rpf$ background is determined by Monte Carlo simulation of $\jpsi \ar \rpf$ and
normalized to the estimated number of $\jpsi
\ar \rpf$ background events.  The fitting result
is shown in Fig. \ref{mpi1pi4fit}. The $\mpp$ spectrum is
the sum of the $M_{\pi 1\pi 4}$ and the $M_{\pi 2\pi 3}$
spectra. The fitted number of
$f_{2}$ component in the $\mpp$ spectrum is 1292 (both
projections are used).

\begin{figure}[htbp]
\centerline{ \hbox{\psfig{file=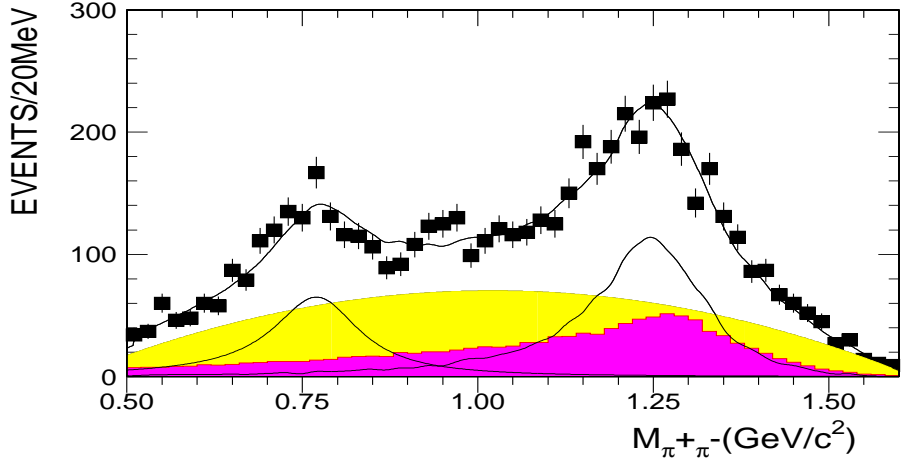, width=13cm,
height=6.5cm}} } \caption{Fit of the $\mpp$
spectrum to obtain the $f_{2}$ signal. The light shaded area is the 2nd order polynomial
background. The dark shaded area is $\rpf$ background. The 
two peaks are $\rho^{0}$ and $f_{2}(1270)$, respectively. }
\label{mpi1pi4fit}
\end{figure}

The branching ratio of $\jpsi \ar \gff$ can be determined with
the following formula:\\
\begin{align}
   &  \nonumber \\
Br(\jpsi\ar \gff) & = \frac{N^{obs}/\varepsilon}{N_{\jpsi}\cdot Br^{2}_{f_{2}\ar\pp}} \nonumber \\
 & = (9.5\pm0.7)\times 10^{-4} \nonumber
\end{align}
where $N^{obs}$=646(=1292/2) is the average number of survived $\jpsi \ar \gff$
events in $M_{\pi 1\pi 4}$ projection and $M_{\pi 2\pi 3}$ projection, 
$\varepsilon=3.71\%$ is the
efficiency determined from phase space
 Monte Carlo events with the same cuts used to select
 $\jpsi\ar\gff$ data sample,
 $N_{\jpsi}=57.7\times 10^{6}$ is the total
number of $\jpsi$ events collected by BES\,II
~\cite{fangss}, and $Br_{f_{2}\ar\pp}=0.565$ is the
branching ratio of $f_{2}\ar \pp$~\cite{pdg}.

\begin{table}[htpb]
\caption{Summary of systematic errors }
\begin{center}
\begin{tabular}{|l|c|}
  \hline
  Sources & Systematic error ($\%$) \\ \hline
  MDC tracking efficiency & 8 \\
  PID efficiency & 3 \\
  Photon selection & 2 \\
  Kinematic fit & 4 \\
  The number of $\jpsi$ events  & 4.7 \\
  $f_{2}$ width & 5.1 \\
  $\mpppp$ Cut & 9.7 \\
  Background uncertainty& 5.5 \\
  $Br_{f_{2}\ar\pp}$ & 2.8 \\ \hline
  Total & 16.5 \\ \hline
\end{tabular}
\label{syserr}
\end{center}
\end{table}

Many sources of systematic errors are considered. Systematic
errors associated with efficiencies, such as from the MDC 
tracking, particle identification,
photon selection, and kinematic fitting, are determined by
comparing $\psi(2S)$ and $\jpsi$ data with Monte Carlo simulation for very
clean decay channels, such as $\psi(2S) \ar
\pi^+\pi^-\jpsi$.

The MDC tracking efficiency has been measured using channels
like $\jpsi\ar \Lambda\overline{\Lambda}$ and $\psi(2S) \ar
\pi^+\pi^-\jpsi$, $\jpsi \ar \mu^{+} \mu^{+}$. It is found
that the efficiency of the Monte Carlo simulation agrees
with that of data within 1-2\% per charged track. The
total systematic error from the uncertainty of MDC tracking
efficiency in our analysis is
taken as 8\%.
The particle identification (PID) efficiency systematic error is calculated by comparing the
efficiency of data with that of Monte Carlo.  The
largest difference between these two efficiencies is within
3\%.
According to the study of photon detection efficiency in
$\jpsi\ar \rho\pi$ ~{\cite{lismfangss}}, SIMBES
simulates the photon detection efficiency in the full energy
range within 1 to 3\%. Here, we take 2\% as the
systematic error in the photon detection efficiency.

The systematic error for the kinematic fit is caused by the
differences between data and simulated data in the momenta and the
error matrices of charged tracks and the energies and the directions
of neutral tracks.  To check the consistency between data and
Monte Carlo simulation, two channels, $\jpsi\ar \rho^{0}\pi^{0}$ and
$\jpsi\ar \Lambda\overline{\Lambda}$, are analyzed for 2-prong and
4-prong events respectively. The systematic error for 4-prong events
caused by kinematic fit is determined to be 4\%.

Other sources of systematic errors come from the uncertainty
of $N_{\jpsi}$, the $f_{2}$ width, the $\mpppp$ requirement, and the
uncertainty of the
background. The total $\jpsi$ number is $(57.7\pm
2.7)\times 10^{6}$ determined from 4-prong events ~\cite{fangss}.
The width of the $f_{2}$ is $185.1^{+3.4}_{-2.6}$ MeV ~{\cite{pdg}}; 185.1 MeV is used in
the Monte Carlo simulation. To estimate the systematic error caused by
the width of $f_{2}$, 188.5 MeV and 182.5 MeV are also tried. After refitting the $\pp$ mass spectrum
of Fig. \ref{mpi1pi4fit}, the relative differences of the branching ratios determined by 
using these two widths compared with the original one are 5.1\% and -3.5\% respectively. The systematic
error caused by the $f_{2}$ width is taken as 5.1\%.

The requirement
$\mpppp \ge 2.6{\rm~GeV}$ is used to select events for
the determination of
$Br(\jpsi\ar \gff)$. The $\mpppp$ spectra of real data and
Monte Carlo events are somewhat different. To estimate the
systematic error caused by this requirement, we calculate
$Br(\jpsi\ar \gff)$ with $\mpppp\ge2.7{\rm~GeV}$ and compare
the result with that obtained with $\mpppp\ge2.6{\rm~GeV}$. The
relative difference is 9.7\%, and this is taken as one of the
systematic errors.

The background systematic error is caused by
the uncertainty in the $\jpsi \ar \rpf$ background.
This uncertainty is
determined by varying the selection criteria
used to obtain the number of $\jpsi \ar \rpf$ events.
The systematic error caused by the uncertainty of $\jpsi \ar \rpf$ 
background is 5.5\%.


Table \ref{syserr} lists all the systematic errors, as well as the
total systematic error of 16.5\%, obtained by adding all
contributions in quadrature. The branching ratio of $\jpsi\ar \gfmfm$
is $$Br(\jpsi\ar \gfmfm) = (9.5\pm0.7\pm1.6)\times 10^{-4},$$
where the first error is statistical and the second
systematic.  This branching ratio includes contributions from
intermediate states such as $J/\psi \ar \gamma \eta_c$.

\section{Summary}
A new decay mode $\jpsi\ar \gfmfm \ar \gpppp$ is observed,
and its branching ratio is measured to be
$Br(\jpsi\ar\gff)=(9.5\pm0.7\pm1.6)\times 10^{-4}$. This result will be helpful to understand the complex
$\jpsi\ar \gpppp$ decay.  


\section{Acknowledgment}
\vspace{0.4cm}

   The BES collaboration thanks the staff of BEPC for their
hard efforts.  This work is supported in part by the
National Natural Science Foundation of China under contracts
Nos. 19991480,10225524, 10225525, the Chinese Academy  of
Sciences under contract No. KJ 95T-03, the 100 Talents
Program of CAS  under Contract Nos. U-11, U-24, U-25, and
the Knowledge Innovation Project of  CAS under Contract
Nos. U-602, U-34 (IHEP); by the National Natural Science
Foundation of China under Contract No.10175060 (USTC), and
No.10225522 (Tsinghua University);  and by the Department  of
Energy under Contract No.DE-FG03-94ER40833 (U Hawaii).


\end{document}